\documentclass{rstransa}
\usepackage{graphicx}
\usepackage{titlesec}
\usepackage{cite}
\usepackage{siunitx}
\usepackage{enumitem}
\usepackage[version=3]{mhchem}
\jname{rsta}
\DeclareSIUnit\flux{\watt\per\square\meter}
\DeclareSIUnit\heat{\kelvin\per\day}
\Journal{Phil. Trans. R. Soc}
\begin{document}
	\title{Predicting atmospheric optical properties for radiative transfer computations using neural networks}
	\author{Menno A. Veerman$^{1}$, Robert Pincus$^{2,3}$, Robin Stoffer$^{1}$, Caspar M. van Leeuwen$^{4}$, Damian Podareanu$^{4}$,  Chiel C. van Heerwaarden$^{1}$}
\address{$^{1}$Meteorology and Air Quality Group, Wageningen University and Research, Wageningen, The Netherlands, $^{2}$Cooperative Institute for Research in Environmental Sciences, University of Colorado at Boulder, Boulder, USA, $^{3}$NOAA Physical Sciences Laboratory, Boulder, USA, $^{4}$SURFsara, Amsterdam, the Netherlands}
\subject{Atmospheric science, Meteorology, Artificial intelligence}  \corres{Menno A. Veerman\\ \email{menno.veerman@wur.nl}}

\begin{abstract}
The radiative transfer equations are well-known, but radiation parametrizations in atmospheric models are computationally expensive. A promising tool for accelerating parametrizations is the use of machine learning techniques. In this study, we develop a machine learning-based parametrization for the gaseous optical properties by training neural networks to emulate a modern radiation parameterization (RRTMGP).
To minimize computational costs, we reduce the range of atmospheric conditions for which the neural networks are applicable and use machine-specific optimised BLAS functions to accelerate matrix computations. To generate training data, we use a set of randomly perturbed atmospheric profiles and calculate optical properties using RRTMGP. Predicted optical properties are highly accurate and the resulting radiative fluxes have average errors within \SI{0.5}{\flux} compared to RRTMGP. Our neural network-based gas optics parametrization is up to 4 times faster than RRTMGP, depending on the size of the neural networks. We further test the trade-off between speed and accuracy by training neural networks for the narrow range of atmospheric conditions of a single large-eddy simulation, so smaller and therefore faster networks can achieve a desired accuracy. We conclude that our machine learning-based parametrization can speed-up radiative transfer computations whilst retaining high accuracy.  \end{abstract}
\maketitle
	\section{Introduction}
	Accurate calculations of radiative fluxes are key to capturing the coupling between radiation, the atmosphere, and the surface. 
	Unlike many parametrizations of subgrid processes, the equations governing radiative transfer are well-known.
	this integration is often parameterized with correlated-$k$ distribution methods \cite{Goody1989,Lacis1991,Fu1992} to drastically reduce the number of quadrature points. 
	Even with this approximation radiative transfer schemes in weather and climate models remain a large computational burden. 
	An important part of radiative transfer parametrizations is therefore to find approaches or approximations that further reduce the computational costs, for example by coarsening the spatial and temporal resolution of the radiative transfer computations \cite{Morcrette2000,Hogan2018} or
	random sampling in spectral space \cite{Pincus2009}.

	A promising and increasingly explored approach to accelerate or improve parametrizations is the use of machine-learning techniques \cite{Reichstein2019}. 
	The application of machine learning to accelerate expensive radiative transfer computations was one of the first uses of machine learning in forward modelling in the atmospheric sciences \cite{Chevallier1998}, and a range of studies have used machine learning to predict vertical profiles of longwave \cite{Chevallier1998,Krasnopolsky2005,krasnopolsky2006,Krasnopolsky2010,Belochitsky2011,Pal2019} and shortwave \cite{krasnopolsky2006,Krasnopolsky2010,Pal2019} radiative fluxes in weather and climate models. 
	These end-to-end approaches, i.e. predicting radiative fluxes by fully emulating a radiative transfer scheme, may result in speed-ups of more than one order of magnitude \cite{Chevallier1998,Krasnopolsky2005,Krasnopolsky2010}, with root mean squared errors of the heating rates within \SI{0.2}{\heat} for shortwave and \SI{0.5}{\heat} for longwave radiation. 
	However, such an approach is inflexible with respect to changes in model configuration including the vertical discretization as the neural networks are trained for a fixed number of vertical layers \cite{Chevallier1998,Krasnopolsky2005,krasnopolsky2006,Krasnopolsky2010}. 
	Moreover, this approach does not respect the well-understood underlying physics.  
	
	In this study we describe a machine-learning approach for accelerating radiative transfer computations that respects the well-understood governing equations, using machine learning only to accelerate the data-driven aspects of the calculations. 
	We exploit the fact that radiation calculations are composed for two distinct steps: the mapping of the physical and chemical state of the atmosphere to a problem in radiative transfer and the subsequent solution of that radiative transfer problem.
	The first step converts temperature, pressure, and composition into atmospheric optical properties that determine how much radiation is emitted (Planck source function), absorbed or scattered (optical depth, single scattering albedo), and the direction of scattering (asymmetry parameter). 
	Unlike the solution of the radiative transfer equation, the calculation of gas optics relies on empiricism and large amounts of tabulated data. 
	
	Here we use machine learning to emulate the calculation of gaseous optical properties of a modern  RRTM for General circulation model applications - Parallel (RRTMGP) \cite{Pincus2019}. 
	We use neural networks as a computationally efficient tool to replace the lookup-tables used in RRTMGP and train the networks for a narrowed range of atmospheric conditions, e.g. by neglecting variations of many trace gases and limiting the range of temperatures and pressures. 
	Constraining the range and number of inputs allows further optimisation of the neural networks, which helps to reduce the computational costs of our parametrization compared to RRTMGP. 
	This approximation can be particularly suitable for limited-area models, such as large-eddy simulations (LES), in which the range of values of thermodynamic variables is smaller than is required by general-purpose tools.  

	\section{Training data generation}\label{sect:datagen}
	We train three sets of artificial neural networks to predict all optical properties: 
	\begin{itemize}
		\itemsep0em 		
		\item One set (\texttt{NWP}) is trained for a wide range of atmospheric conditions, roughly representing the variability expected in global numerical weather prediction, but with all gases except water vapour and ozone kept constant;
		\item Two sets (\texttt{Cabauw}, \texttt{RCEMIP}) are trained for only the narrow range of atmospheric conditions within one LES simulation each, in order to estimate the performance gains of this LES-specific tuning. 
	\end{itemize}
    Because the LES-tuned networks are trained for a narrow range of atmospheric conditions, they may contain substantially fewer layers and nodes than the NWP-tuned networks and can therefore be faster. 
    This may provide a great speed-up when an LES simulation is repeated for large number of numerical experiments.

	To generate the data required to train and validate the neural networks we use the new radiation package RTE+RRTMGP \cite{Pincus2019}. 
	RRTMGP is a parameterization of gas optics, computing all optical properties from temperature ($T$), pressure ($p$), and the concentrations of a wide range of gases. 
	Given these optical properties, RTE (Radiative Transfer for Energetics) calculates the radiative fluxes throughout each column. 
	
	RRTMGP covers the spectral range of radiation relevant to atmospheric problems using a correlated \textit{k}-distribution \cite{Fu1992} with 14 shortwave (\SI{0.2}{\micro\meter}-\SI{12}{\micro\meter}) bands, 16 longwave (\SI{3}{\micro\meter}-\SI{1000}{\micro\meter}) bands, and 16 \textit{g}-points per band. 
	We therefore need to predict 224 ($14\times16$) values for the short wave optical properties and 256 ($16\times16$) for the longwave optical depth. 
	Additional to calculating the Planck source functions of each layer from the layer temperature, RRTMGP also calculates the Planck source functions at the layer interfaces from the interface temperatures and the wavelength to $g$-point mapping of the layers below and above each interface, separately. 
	Each layer interface thus has two Planck source functions, representing the upward and downward emission. 
	Since we aim to predict the same output as RRTMGP, we also need to predict the upward and downward emission at each layer interface, resulting in 768 ($3\times16\times16$) values per grid cell.. 
	
	The NWP-tuned neural networks are trained with only $T$, $p$, \ce{H2O} and \ce{O3} as input, which are time-varying 3D variables in global weather prediction models. To create training data, we begin with the set of 100 atmospheric profiles of temperature, pressure and several gas concentrations from the Radiative Forcing Model Intercomparison Project (RFMIP) \cite{Pincus2016}.
	These 100 profiles were chosen to reproduce the global, annual mean radiative forcing between present-day and pre-industrial conditions \cite{Pincus2016}, but do not represent the full diversity of atmospheric conditions on earth. RFMIP assumes that all gases except \ce{H2O} and \ce{O3} are well-mixed and we train only on present-day concentrations to reduce the degrees of freedom. Since larger amounts of training data is required to train neural networks, we generate extra data, spanning a slightly wider range of atmospheric conditions, by generating permutations of this set of 100 profiles with random perturbations in $T$, $p$, \ce{H2O} and \ce{O3}:
	$$H_2O^i=(1+c_1r_1)H_2O^i,\ O_3^i=(1+c_2r_2)O_3,\ T^i=T^i+c_3r_3,\ p^i=(p^{i+1}-p^{i-1})r_4+p^{i-1},$$
	where $r_1$, $r_2$ and $r_3$ are random numbers between $-1$ and $1$, $r_4$ is a random number between 0.05 and 0.95, and $i$ is the index of each layer. The choice of constants $c_1$, $c_2$ and $c_3$, which define the maximum width of the permutations, determines the range of atmospherics conditions the networks are trained for, but also affect the risk of generating unrealistic data. Here, we set constants $c_1$, $c_2$ and $c_3$ to $\frac{3}{4}$, $\frac{3}{4}$ and 5, respectively. Since the four random numbers are generated independently, the perturbations are uncorrelated. This gives a larger variation of different atmospheric conditions in the training data and reduces the risk of overfitting, but may result in combinations of $T$, $p$, \ce{H2O} and \ce{O3} that are unlikely to occur in reality. Water vapor mixing ratio is constrained to be less than saturation. 
	By generating $r_4$ between 0.05 and 0.95 and not between 0 and 1, we aim to prevent unrealistically small layer depths.
	To obtain a wider range of surface conditions, we randomly choose the surface temperature between $T_0-$\SI{10}{\kelvin} and $T_0+$\SI{10}{\kelvin}, where $T_0$ is the temperature at the lowest pressure level. Note that the surface temperature is not used for training, since we compute the surface emission from the emission of the lowest model layer (Eq. \ref{eq:eq3}), but only to compute radiative fluxes from the predicted optical properties.
	
	For the two LES-tuned sets of neural networks, which are trained for a narrow range of atmospheric conditions, we compute \ce{O3} as a monotonic function of pressure following \cite{Wing2018} with a lower boundary of \SI{5e-3}{ppmv}. Consequently, we train these networks using only $T$, $p$ and \ce{H2O}. For the \texttt{Cabauw} networks, we run a 10-hour LES simulation (07 UTC to 17 UTC) of a developing convective boundary layer over grassland near the Cabauw Experimental Site for Atmospheric Research (CESAR) in the Netherlands with shallow cumulus clouds forming in the afternoon (see \cite{Vila2014} and \cite{Pedruzo17} for a detailed description of the case). The simulation is performed using the Dutch Atmospheric Large-Eddy Simulation (DALES) \cite{Heus2010}, with domain size of \SI[product-units=single]{19.2x19.2x5.47}{\kilo\meter^3} and a resolution of \SI[product-units=single]{100x100x24}{\meter^3}. For \texttt{RCEMIP} networks, we run a 100-day simulation with MicroHH \cite{Heerwaarden2017} following the specification for cloud resolving models (see \texttt{RCE\_small300} in \cite{Wing2018}) of the Radiative Convective Equilibrium Model Intercomparison Project  (RCEMIP) \cite{Wing2018}. This is a case with deep convection over a tropical ocean with an atmosphere in radiative convective equilibrium, meaning that radiative cooling is balanced by convective heating \cite{Wing2018}. The simulation is performed with a domain size of \SI[product-units=single]{100x100x32}{\kilo\meter^3}, a horizontal resolution of \SI{1}{\kilo\meter} and 72 vertical levels. For each LES-tuned set, we then determine the minimum and maximum \ce{H2O} and $T$ values of each vertical layer and subsequently generate random profiles of $p$, \ce{H2O} and $T$ to cover the full parameter space of the corresponding simulation. We generate 1000 profiles for the \texttt{Cabauw} networks and 3000 profiles for the \texttt{RCEMIP} networks, because the RCEMIP simulation has a higher domain top and therefore spans larger range of atmospheric conditions. To deal with negative or unrealistically low water vapour concentrations in the simulations, we set a lower \ce{H2O} limit in these profiles of \SI{16}{ppmv} and \SI{5}{ppmv} for the \texttt{Cabauw} and \texttt{RCEMIP} simulation, respectively. Since the LES-tuned networks are trained for a narrow range of atmospheric conditions, they need to be retrained to be used for a LES simulation with a different range of conditions, which takes on the order of 1 hour on a single compute node.
	
	For every combination of $T$, $p$, \ce{O3} and \ce{H2O} (\texttt{NWP}) or $T$, $p$, and \ce{H2O} (\texttt{Cabauw}, \texttt{RCEMIP}), we then calculate the optical properties at each $g$-point using RRTMGP. 
	During both training and inference, we log-scale the optical depths, Planck source function, \ce{H2O}, (\ce{O3},) and $p$, which improves the convergence of the neural networks because the range of these variables may span multiple orders of magnitude. 
	For the log-scaling, we use a fast approximation of the natural logarithm to keep the computational effort feasible: 
	\begin{align}\label{eq:eq1}
    \ln{x}=\lim_{n \to\infty}(\sqrt[n]{x}-1)n	\end{align} where we take $n=16$. This approximation is consistent with the fast approximation of the exponential we use during inference (Eq. \ref{eq:eq3}).

	\section{Artificial neural networks}
	\subsection{Data preparation}
	Before training our neural networks we normalize all variables to a zero mean and unit variance, with different means and standard deviations for the upper ($p<$\SI{9948}{Pa}) and lower ($p>$\SI{9948}{Pa}) atmosphere per variable. 
	A random 95\% of the dataset is used to train the neural networks and 5\% of the data is used for validation. 
	In total, the training datasets for the \texttt{NWP}, \texttt{Cabauw} and \texttt{RCEMIP} networks consist of about $5.7\times10^5$ (lower atmosphere: $3.4\times10^5$, upper atmosphere: $2.3\times10^5$), $2.2\times10^4$ (lower atmosphere only) and $6.8\times10^4$ (lower atmosphere: $3.8\times10^4$, upper atmosphere: $3.0\times10^4$) data points, respectively.

	\subsection{Network architecture and training}
	The neural networks are designed in and trained with TensorFlow \cite{Abadi2016}, version 1.11/1.12. 
	We need to predict 4 different optical properties: the single scattering albedo $w_0$ (shortwave only), the shortwave $\tau_{sw}$ and longwave $\tau_{lw}$ optical depth, and the Planck source function $B$. 
	Because optical depth depends on layer depth, the optical depths are normalised by the layer thickness ($\Delta p$). 
	We therefore predict $\frac{\tau_{sw}}{\Delta p}$ and $\frac{\tau_{lw}}{\Delta p}$, which are the proportional to the absorption coefficients ($\mathrm{Pa^{-1}}$), so we do not have to account for layer thickness.
	For each optical property we train two neural networks, for the upper atmosphere ($p<$\SI{9948}{Pa}) and lower ($p>$\SI{9948}{Pa}) atmosphere, a distinction that also made in RRTMGP. 
	By training separate neural networks for each optical property we can reduce the complexity of each network. 
	Furthermore, this allows us to (re-)train the networks for the different optical properties independently. 
	The neural networks predict the optical properties of each grid cell or layer from the values of  $T$, $p$, (\ce{O3}), and \ce{H2O} of only that grid cell or layer. 
	As such, each network has either 4 (NWP-tuned set) or 3 (LES-tuned sets) inputs and either 224 ($\frac{\tau_{sw}}{\Delta p}$, $w_0$), 256 ($\frac{\tau_{lw}}{\Delta p}$) or 768 ($B$) outputs. 
	
	All networks are feed-forward multi-layer perceptrons with densely-connected layers. We use a leaky ReLu activation function \cite{Maas2013} with a slope of 0.2 in all hidden layers and a linear activation function in the output layer. We train for 500 epochs (one epoch is one iteration over the entire training dataset). During training, we use the Adam optimiser \cite{Kingma2014} to optimise the weights, with a batch size of 128 and an initial learning rate of 0.01 that decays every 10 epochs. As the loss function, we use the mean squared error (MSE):
	\begin{align}MSE = \frac{1}{N_{batch}N_{gpt}}\sum_m^{N_{batch}}\sum_n^{N_{gpt}}(NN_{m,n}-RR_{m,n})^2,\end{align} 
	where $N_{batch}$ and $N_{gpt}$ are the batch size and number of $g$-points, respectively. $NN$ and $RR$ are the optical properties predicted by the neural networks and calculated by RRTMGP, respectively. 
	
	Although several studies have already shown successful attempts to optimise neural network architecture for accuracy with machine learning \cite{Baker2016,Zoph2016}, choosing the numbers of hidden layers and the number of nodes per layer is often still a matter of manual tuning.
	Wider and deeper networks are able to learn more complex functions, with the risk of overfitting, but are slower during both training and inference. 
	We test several network sizes (Table \ref{tbl:t1}) to investigate the trade off between accuracy and performance for different network sizes. 
	We also tested a network without hidden layers that performs only linear regression as a reference. 
	However, the prediction errors of the linear networks were over 2 order of magnitude higher than the errors of the other networks (not shown), suggesting that a linear parametrization cannot capture the complex relationship between atmospheric conditions and optical properties. 

\begin{table}[]
	\centering
	\caption{Properties of hidden layers, number of nodes, and number of weights (including biases) for all tested neural networks}
	\begin{tabular}{|l|l|l|l|l|}
		\hline
		Name & Layer 1 & \multicolumn{1}{r|}{Layer 2} & Layer 3 & \# of weights\\ \hline \hline
		1L$-$32          & 32      & -                            & - &  8608    \\ \hline
		1L$-$64          & 64      & -                            & - & 35232     \\ \hline
		2L$-$32\_32      & 32      & 32                           & - & 18272     \\ \hline
		2L$-$64\_64      & 64      & 64                           & - & 73312     \\ \hline
		3L$-$32\_64\_128 & 32      & 64                           & 128 & 116928   \\ \hline
	\end{tabular}
	\label{tbl:t1}
\end{table}

\subsection{Implementation}{\label{sect:impradsolve}}
We use the trained neural networks as a parametrization for the gas optics in the RTE+RRTMGP framework. 
We replace the RRTMGP gas optics and source function routine and pair the new parametrization with the RTE radiative transfer solver. 
The new parametrization gets as input one or more columns of $T$, $p$, \ce{H2O} and \ce{O3} (latter only for \texttt{NWP}) and outputs, for each layer of each column, the log-scaled and normalised optical properties for all 224, 256 or 768 $g$-points. 
The workflow for a neural network with two hidden layers of 64 nodes, is as follows:
\begin{enumerate}
	\item Initialise the weights matrices $W_1[N_1,N_{in}]$, $W_2[N_2,N_1]$, $W_3[N_{out},N_2]$ and bias vectors $\beta_1[N_1]$, $\beta_2[N_2]$, $\beta_3[N_{out}]$ of the trained neural networks, where $N_{in}$ is the number of inputs (3 or 4), $N_{out}$ the number of outputs (224, 256 or 768), and $N_1$ and $N_2$ the number of nodes of the first and second hidden layer (both 64).
	\item Create input matrix $I[N_{in},N_{batch}]$ from the 4 input columns, where $N_{batch}$ is the batch size, i.e. the number of grid cells computed simultaneously, where the number of grid cells is the number of atmospheric profiles multiplied by the amount of vertical layers per profile.
	\item Apply log-scaling on $p$, \ce{H2O} and \ce{O3} (\texttt{NWP} only) and normalise all input variables
	\item Calculate the first hidden layer ($L_1[N_1,N_{batch}]$):
	\begin{itemize}
		\item[a.] Calculate matrix product $L_1=W_1I$ 
		\item[b.] Add $\beta_1$ to each column of $L_1$
		\item[c.] Apply Leaky ReLu activation function
	\end{itemize}
	\item Calculate the second hidden layer ($L_2[N_2,N_{batch}]$):
	\begin{itemize}
		\item[a.] Calculate matrix product $L_2=W_2L_1$ 
		\item[b.] Add $\beta_2$ to each column of $L_2$
		\item[c.] Apply Leaky ReLu activation function
	\end{itemize}	
	\item Calculate output matrix ($O[N_{out},N_{batch}]$):
	\begin{itemize}
		\item[a.] Calculate matrix product $O=W_3L_2$ 
		\item[b.] Add $\beta_3$ to each column of $O$
	\end{itemize}
	\item Denormalise output matrix $O$, take the exponential of all values and multiply $\frac{\tau_{sw}}{\Delta p}$ $\frac{\tau_{lw}}{\Delta p}$ with layer thickness $\Delta p$. 
\end{enumerate}
The matrix products are the computationally most expensive parts of the neural network-solver. We therefore make use of the Level 3 functions of the Basic Linear Algebra Subprograms (BLAS) library for which machine-specific optimised versions exist. In our implementation, we use the C-interface to the \texttt{sgemm} function of Intel's Math Kernel Library\cite{Intel} (MKL). For the exponentiation, we use a fast approximation consistent with the natural logarithm approximation (eq. \ref{eq:eq1})
\begin{align}\label{eq:eq3}
e^x=\lim_{n\to\infty}(1+\frac{x}{n})^n,
\end{align} where we take $n=16$. 
The exponential in step vii is omitted for the single scattering albedo $w_0$. The Planck source function at the surface $B_{sfc}$ is calculated from the Planck source function of the lowest layer $B_{lay0}$ with a function of the form \begin{align}\label{eq:eq3}
B_{sfc}=B_{lay0}\left(\frac{\alpha T_{sfc}}{\alpha T_{lay0}}\right)^\beta, 
\end{align}
where $T_{sfc}$ and $T_{lay0}$ are the temperature of the surface and of the lowest layer, respectively, and coefficients $\alpha$ and $\beta$ are fitted (per spectral band) from the lookup tables of RRTMGP. 

\section{Results \& Discussion}

\subsection{NWP-tuned networks}
\subsubsection{Prediction skill}
\begin{figure}[!h]
	\centering
	\includegraphics[width=1.\linewidth]{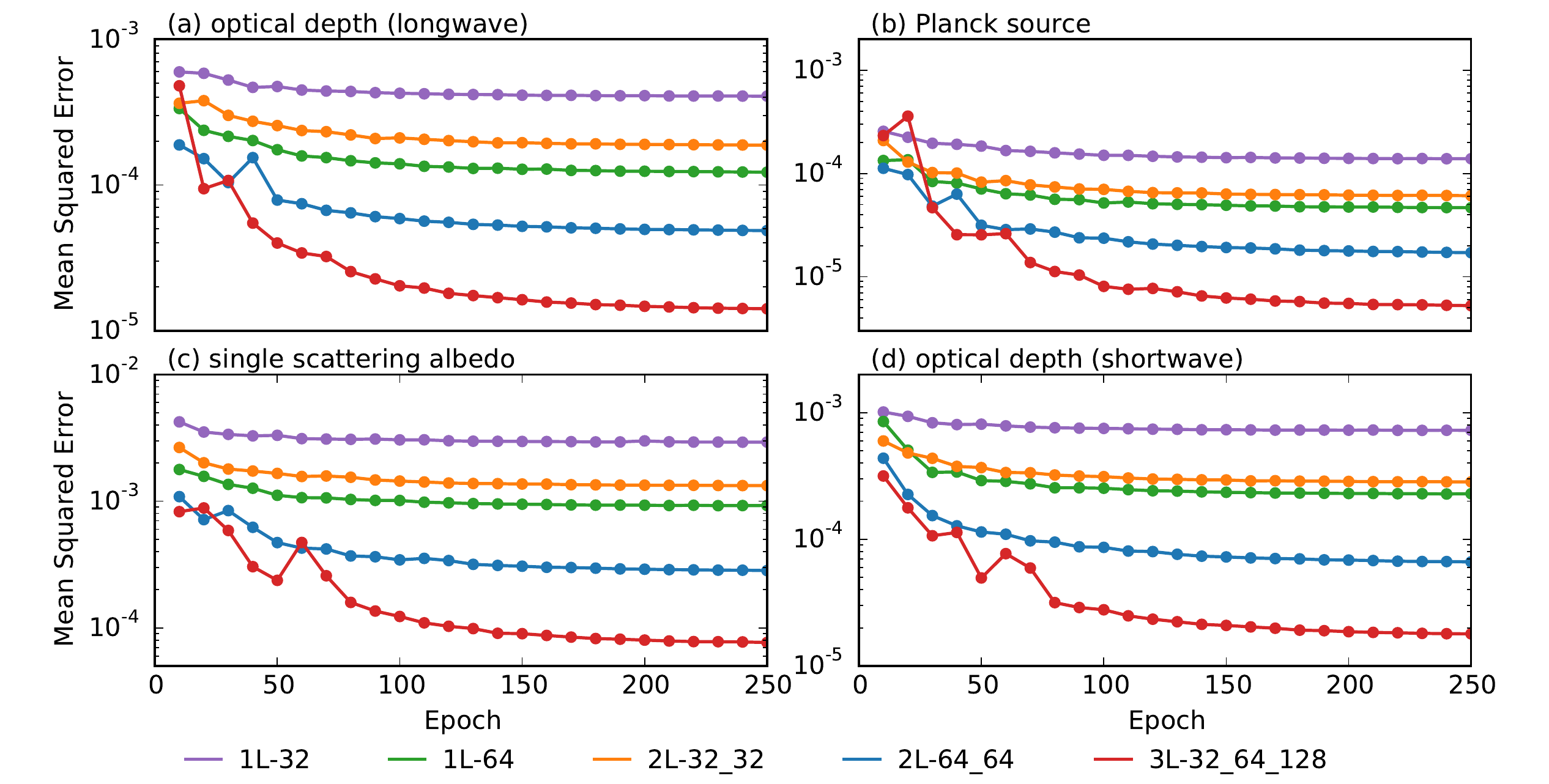}
	\caption{Mean squared errors, as function of the number of epochs, of the longwave optical depth (a), Planck source function (b), shortwave optical depth (c) and single scattering albedo (d) for the six different network sizes (Table \ref{tbl:t1}). The mean squared errors are based on the full validation dataset and the validation is performed every 10 epochs.}
	\label{fig:f1}
\end{figure}

Since our loss function is the mean squared error (MSE) of the predicted optical properties with respect to the optical properties calculated by RRTMGP, a useful indication of the accuracy of the neural networks is the evolution of the MSE as training progresses (Fig. \ref{fig:f1}). We generally see that as the size of the networks increases, the number of epochs needed to reach convergence increases and the MSE at the end of the training decreases. The relatively large difference between the MSE's of the 2L-64\_64 and 3L-32\_64\_128 networks suggests that we may still be able to strongly reduce the MSE by increasing the network size (at the cost of additional computational cost). The 1L\_64 networks are slightly more accurate than the 2L\_32\_32 networks despite having the same number of hidden nodes: due to the larger number of output nodes (224/256), the 1L\_64 networks have more connections between nodes (Table \ref{tbl:t1}). This means that more complex functions can be learned, but may also lead to higher computational costs.

To test whether the neural network predictions correlate well with RRTMGP, we generate a new set of 100 randomly perturbed profiles and calculate R-squared values ($R^2$) between the optical properties computed by the neural networks and by RRTMGP. The $R^2$-values are determined for each $g$-point separately and subsequently averaged over all 224 or 256 $g$-points ($\overline{R^2}$) to represent the overall performance. The networks typically have $\overline{R^2} > 0.9998$ for the optical depths and the Planck source function and $\overline{R^2} > 0.998$ for the single scattering albedo. These high correlations give us confidence that the neural networks are able to predict the optical properties with very high accuracy. 

\subsubsection{Radiative fluxes}{\label{sect:toradflux}}
\begin{figure}[!h]
	\centering
	\includegraphics[width=1.\linewidth]{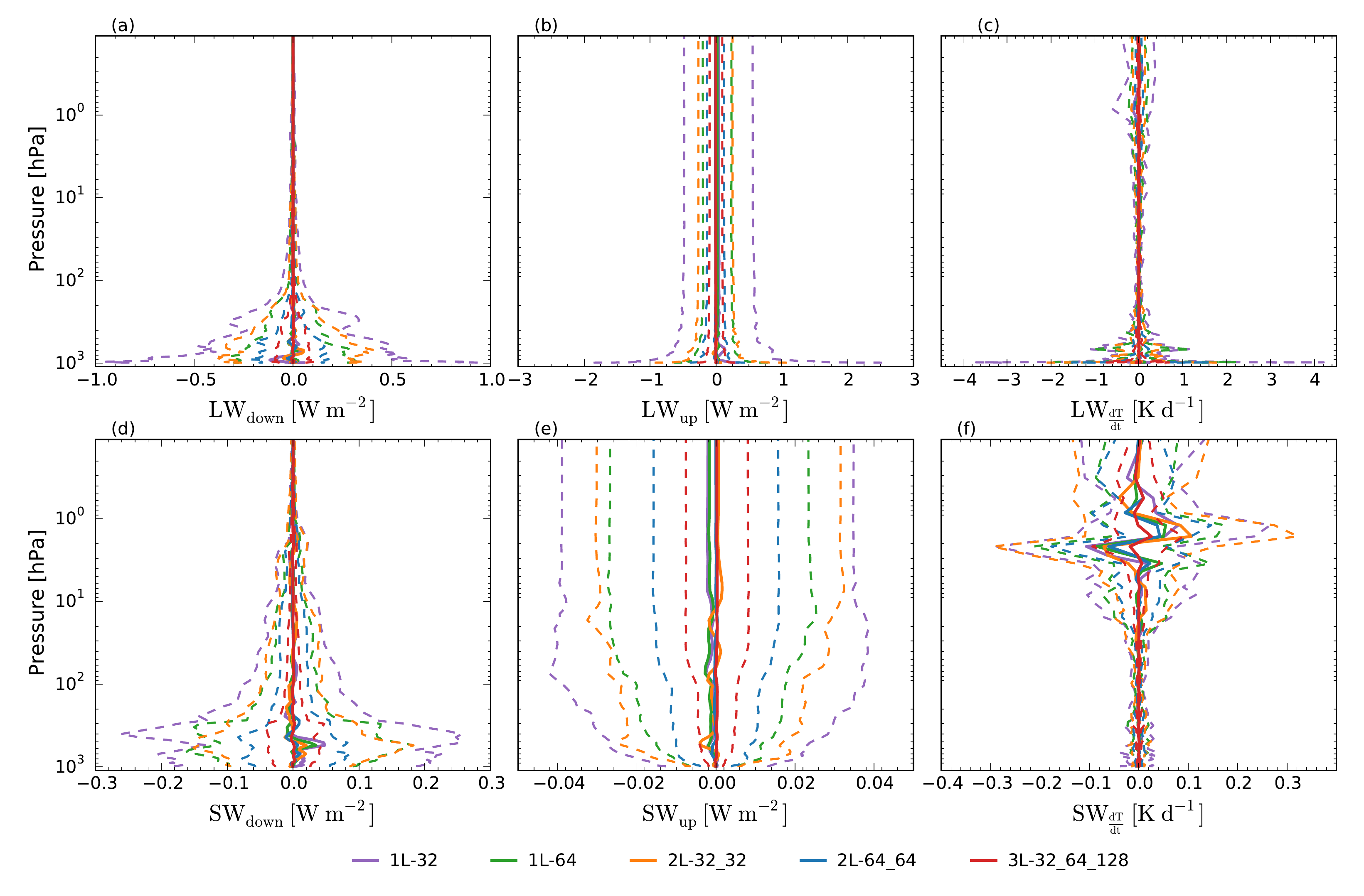}
	\caption{For all network sizes, vertical profiles of the errors of the radiative fluxes and heating rates based on the neural network-predicted optical properties with respect to the radiative fluxes based on the optical properties from RRTMGP. Shown are the mean (solid) and twice the standard deviation (dashed) of the error, for the downwelling (a) and upwelling (b) longwave radiation, longwave heating rates (c), the downwelling (d) and upwelling (e) shortwave radiation and shortwave heating rates (f). A zenith angle of 42$^\circ$ and an albedo of 0.07 is used for the shortwave fluxes.}
	\label{fig:f2}
\end{figure}

The neural networks predict the optical properties very accurately, but for atmospheric modelling applications we require accurate radiative fluxes through the atmosphere and at the surface. To assess whether the accuracy of the neural networks is sufficient, we use our implementation of the neural networks in RTE+RRTMGP to calculate radiative fluxes based on the optical properties predicted by the neural networks.

The errors of the radiative fluxes based on the neural network-predicted optical properties generally decrease as the complexity of the neural networks increases (Fig. \ref{fig:f2}). 
Mean flux errors differ only slightly, whereas the spread of the flux errors is clearly smaller for the more complex networks. 
This indicates that reducing the size of the neural networks used to predict optical properties does not introduce a significant bias in the radiative fluxes, but results in larger flux errors for individual atmospheric profiles because smaller networks predict less accurate optical properties. 

Our average errors in the downwelling surface fluxes and upwelling top of atmosphere fluxes with respect to RRTMGP (Figure \ref{fig:f2}) are similar to or smaller than the average errors of RRTMGP with respect to the line-by-line radiative transfer model (LBLRTM) \cite{Clough2005}, as reported by \cite{Pincus2019} for the original set of RFMIP profiles.
However, the accuracy of our neural network-approach may be lower for individual atmospheric profiles, especially with the smallest networks.
Since the neural networks are trained against RRTMGP, the similar mean errors suggest that further increasing the predictive skill of the networks will not improve the radiative fluxes much on average, as the flux errors compared to the "ground truth" will then be dominated by the errors of RRTMGP with respect to LBLRTM. 

The errors of shortwave radiative heating rates, which are proportional to the divergence of the radiative fluxes, are within about \SI{0.05}{\heat} in troposphere and most of the stratosphere, which is similar to the accuracy of RRTMGP with respect to LBLRTM \cite{Pincus2019}. However, the shortwave heating rates errors of the smallest networks are over \SI{0.05}{\heat} near the top of the atmosphere, which is larger than the error of RRTMGP with respect to LBLRTM. The longwave radiative heating rates have errors within \SI{1}{\heat} for most of the profile, which is higher than the error of RRTMGP with respect to LBLRTM, and the smaller networks give significantly larger errors at the surface.
The largest longwave heating rates errors presumably occur mostly where the temperature gradients of the perturbed profiles are very large, in which case we find that the absolute heating rates of RRTMGP may be over \SI{e2}{\heat}.

\begin{figure}[!h]
	\centering
	\includegraphics[width=1.\linewidth]{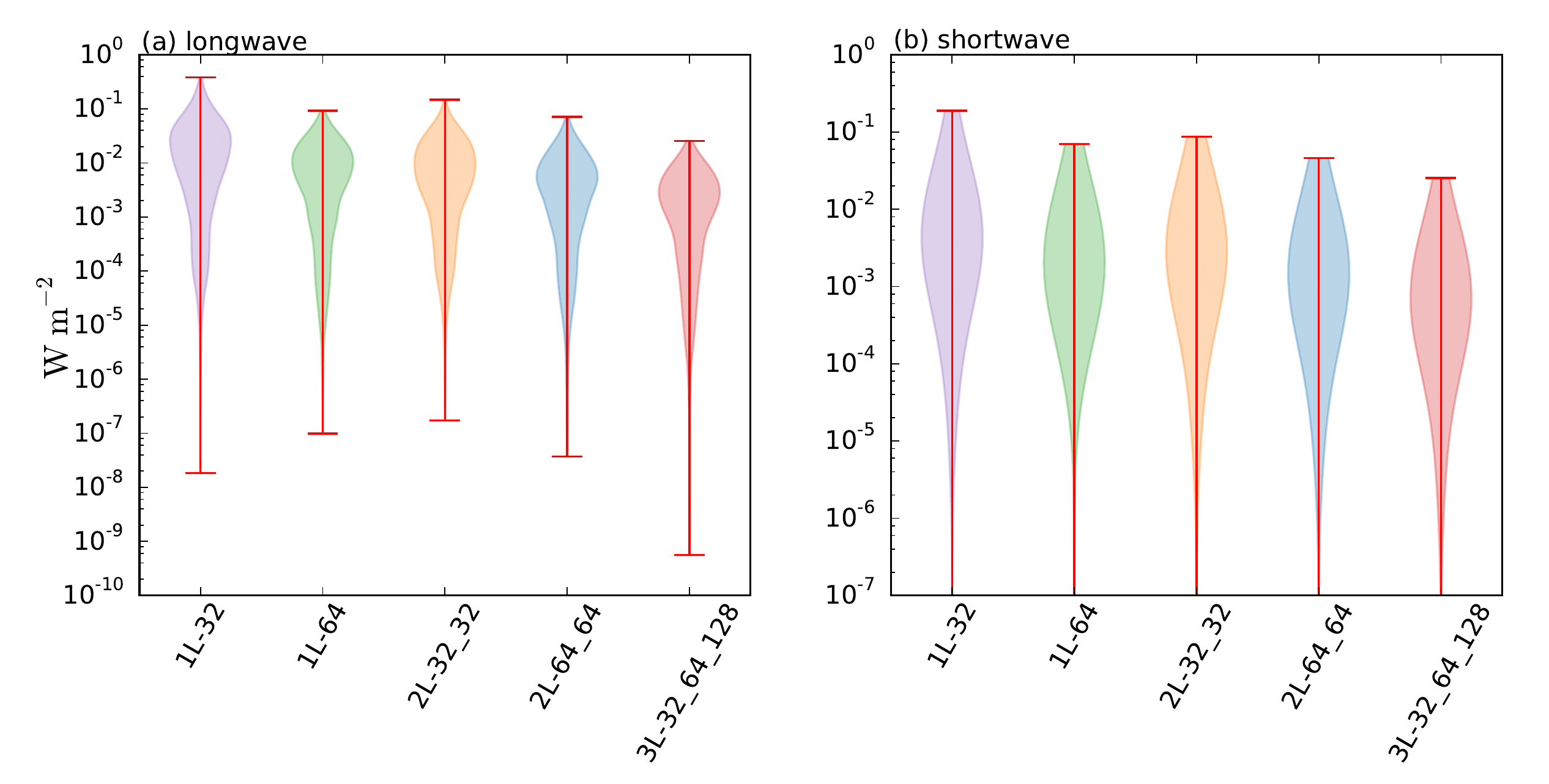}
	\caption{For both longwave (a) and shortwave radiation (b) and for all network sizes, violin plots of the absolute errors per spectral band of the downwelling surface fluxes with respect to RRTMGP. The radiative flux per spectral band is the integral of the radiative fluxes over the 16 $g$-points of each band. For each network size, the minimum errors in (b) are on the order of $10^{-13}$ and correspond to the spectral band with the shortest wavelength, which is almost completely absorbed in the stratosphere.}
	\label{fig:f3}
\end{figure}

Additionally, we determine the absolute errors of the downwelling radiative surface fluxes per spectral band, which help to assess to what extent a single band contributes to the broadband flux errors. The accuracy of the radiative fluxes per band may also be of interest for predictions of UV-index \cite{WHO2002} or photosynthetically activate radiation (PAR). The maximum errors of the flux per band range between \SI{0.026}{\flux} and \SI{0.39}{\flux} in the longwave spectrum and between \SI{0.025}{\flux} and \SI{0.18}{\flux} in the shortwave spectrum (Figure \ref{fig:f3}), depending on network size. Using the neural network-predicted optical properties, we can thus calculate the radiative fluxes of each spectral band with errors below \SI{0.5}{\flux}.

\subsubsection{Computational performance}
\begin{figure}[!h]
	\includegraphics[width=1\linewidth]{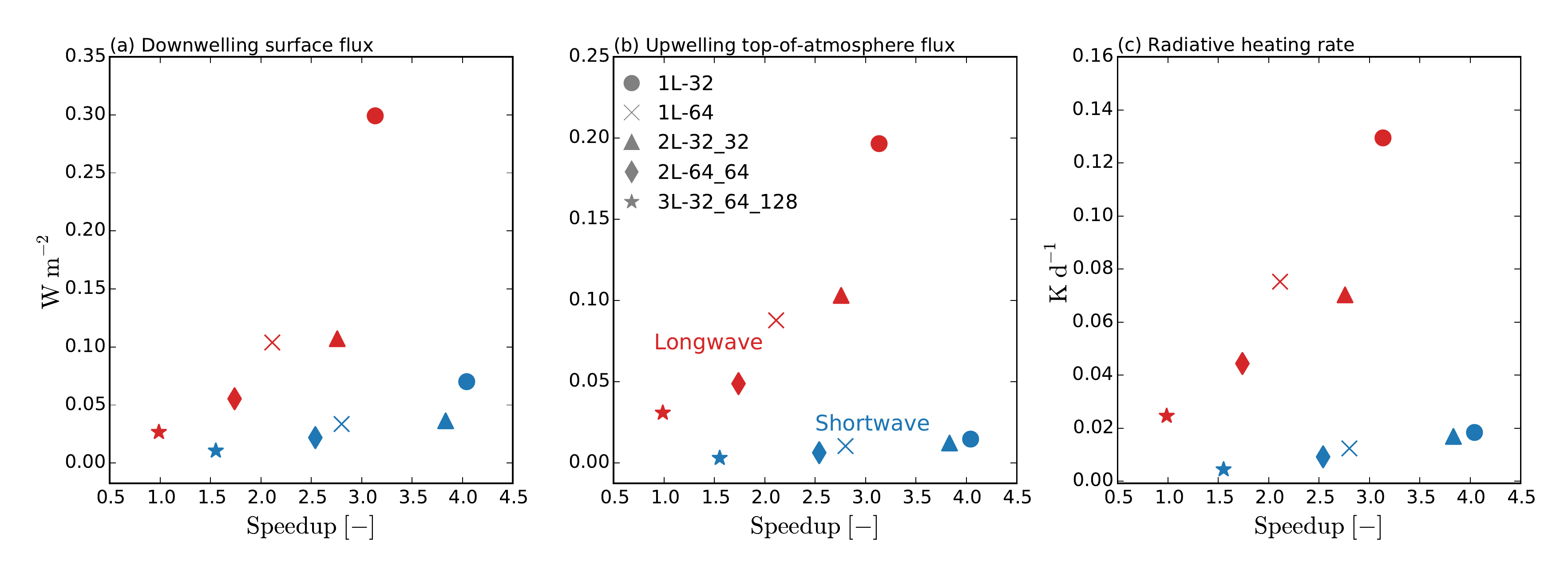}
	\caption{For all network sizes, the speed-up of the neural  networks-solver against the mean absolute errors of the radiative heating rates (a), the upwelling top-of-atmosphere flux (b) and the downwelling surface flux (c), for shortwave (blue) and longwave (red) radiation.}
	\label{fig:f4}
\end{figure}

We have demonstrated that neural networks can accurately reproduce optical properties determined from RRMTGP's lookup tables but the approach is only useful if it can accelerate radiation computations. To evaluate the difference in runtime between RRTMGP and the neural network-solver, we generate 11 additional sets of 100 profiles from the RFMIP profiles. The optical properties of these sets are evaluated sequentially and the runtimes of the last 10 sets are averaged. The runtime of the first set is neglected to allow some spin-up time, mainly for the initialization of BLAS. Since the neural networks run in single precision, we also run RRTMGP in single precision for this benchmark to have a fairer comparison. However, it must be noted that the shortwave solver of RTE currently only works in double precision. The benchmarks are performed on a single core (compute node: $2\times12$-core Intel Xeon E5-2690 v3, 2.6 GHz, 64 GB memory)
 
The neural network-based gas optics parametrization is up to about 4 times faster than RRTMGP, depending on size of the networks (Fig. \ref{fig:f4}). The speed-up of the neural network-solver generally decreases for increasing network complexity. This is as expected, because the number of matrix multiplications scales with the number of layers and because the size of the matrices, and thus the computational effort required for the matrix multiplications, scales with the number of nodes per layer. The choice of a network size thus depends on the acceptable accuracy for a particular range of atmospheric conditions. If the neural networks would be trained for the full range of all 19 gases specified in RFMIP, or for the even larger set of gases treated by RRTMGP, they would have much more input nodes and likely also need more nodes in the hidden layer, which would reduce the speed-up achieved by our neural network approach. 

\subsection{LES-tuned networks}
\begin{figure}[!h]
	\includegraphics[width=1\linewidth]{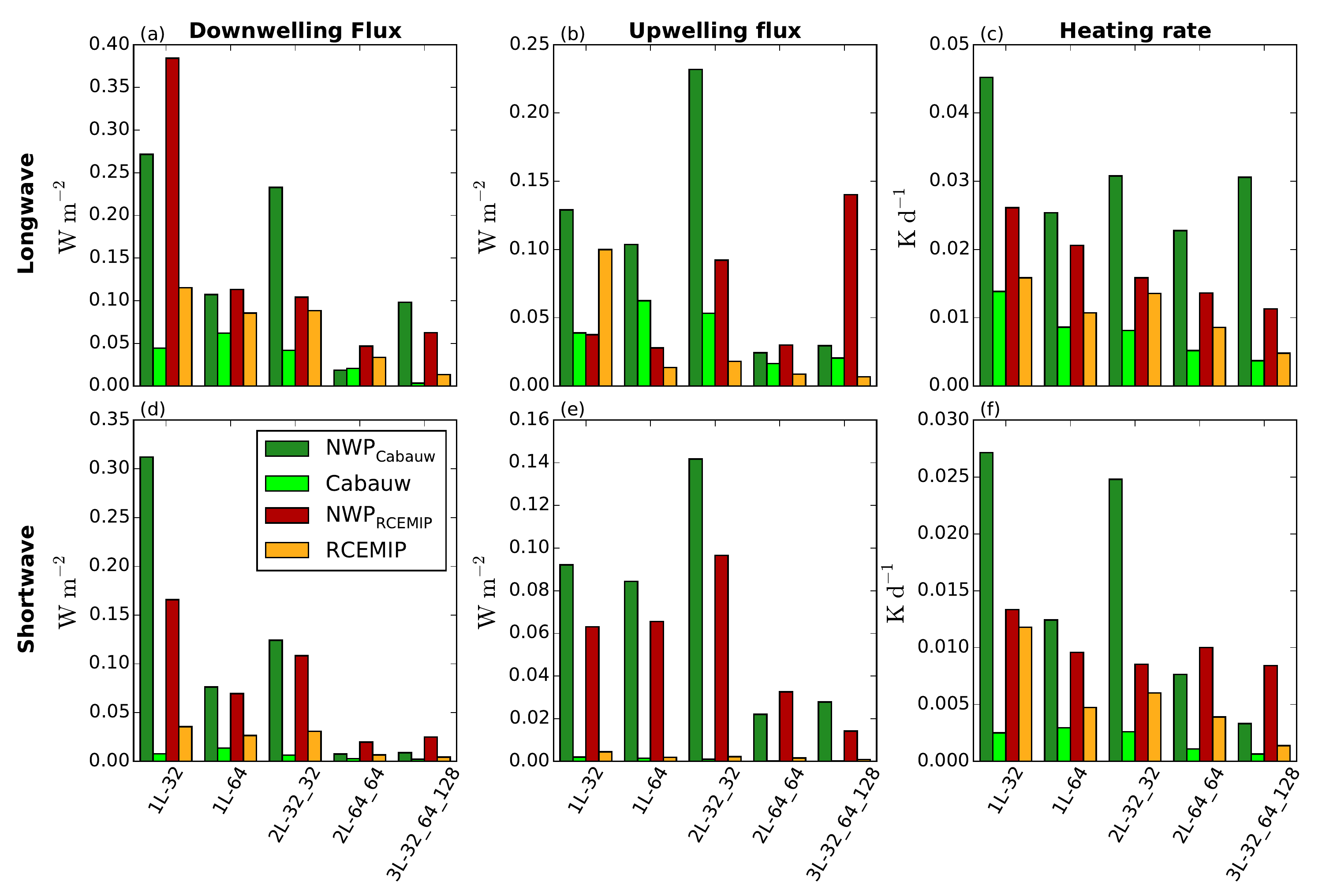}
	\caption{For all network sizes of the \texttt{NWP}, \texttt{Cabauw} and \texttt{RCEMIP} networks, the mean absolute errors with respect to RRTMGP of the radiative heating rates (a,d), upwelling radiative fluxes at the top of atmosphere (b,e) and downwelling radiative fluxes at the surface (c,f) for the longwave (a,b,c) and shortwave (d,e,f) spectrum. Radiative fluxes and heating rates are based on 100 random profiles of the \texttt{Cabauw} simulation (\texttt{$\mathrm{NWP_{Cabauw}}$}, \texttt{Cabauw}) or the \texttt{RCEMIP} simulation (\texttt{$\mathrm{NWP_{RCEMIP}}$}, \texttt{RCEMIP})}
	\label{fig:f5}
\end{figure}

The NWP-tuned networks are trained for a wide range of atmospheric conditions, but in typical LES simulations we may expect only a narrow range of conditions. 
For this reason, we investigate whether smaller and therefore faster network suffice when the networks are trained for the narrow range of atmospheric conditions in a single LES simulation. 
The predicted optical properties of LES-tuned neural networks (\texttt{Cabauw}, \texttt{RCEMIP}) are also very accurate. To further test the LES-tuned neural networks, we sample 100 random atmospheric profiles each from the \texttt{Cabauw} and \texttt{RCEMIP} simulations and calculate profiles of optical properties with the \texttt{Cabauw} and \texttt{RCEMIP} sets of neural networks, respectively. For the \texttt{Cabauw} networks, we generally observe an improvement in the accuracy of the radiative heating rates, the downwelling surface fluxes and the upwelling top-of-atmosphere fluxes (Figure \ref{fig:f5}) due to the LES tuning. This improvement is especially large in the shortwave spectrum, where for some networks sizes the mean squared errors of the surface and top-of-atmosphere fluxes are over an order of magnitude lower with the \texttt{Cabauw} networks than with the \texttt{NWP} networks. The lower mean squared errors of the \texttt{Cabauw} neural networks show that with LES tuning, we can use relatively small networks (e.g. \texttt{1L-32}) to achieve similar or even higher accuracy than the more complex networks (\texttt{2L-64\_64}, \texttt{3L-32\_64\_128}) of the \texttt{NWP} set, which results in a larger speed-up compared to RRTMGP (Figure \ref{fig:f3}). With the \texttt{RCEMIP} neural networks, we also observe a general increase in the accuracy of the shortwave and longwave fluxes and heating rates, (Figure \ref{fig:f5}), but the improvements are often less than with the \texttt{Cabauw} networks. It should be noted that the improvement in accuracy achieved by LES-tuning is not fully consistent. Especially with the smaller networks, the accuracy of the LES-networks may be slightly lower than the accuray of the NWP-tuned networks.

The mean absolute errors of the \texttt{NWP} networks on the profiles of the \texttt{Cabauw} (Figure \ref{fig:f5}) and \texttt{RCEMIP} (Figure \ref{fig:f5}) simulations are frequently larger than the errors of the \texttt{NWP} networks on the RFMIP-based profiles (Figure \ref{fig:f2}). This might be an indication that not all atmospheric conditions occuring in the \texttt{Cabauw} and \text{RCEMIP} simulations are well-enough represented in the training data based on the RFMIP profiles, but the lower errors of \texttt{NWP} networks on the RFMIP-based profiles may also be a sign of overfitting due to insufficiently independent training and testing data. 
Nevertheless, given that the mean absolute errors are well within \SI{0.5}{\flux} we are still confident that the \texttt{NWP} neural networks can be accurately used on a relatively wide range of atmospheric conditions.

\section{Conclusions}
We developed a new parametrization for the gas optics by training multiple neural networks to emulate the gaseous optical properties calculations of RRTMGP \cite{Pincus2019}. The neural networks are able to predict the optical properties with high accuracy and errors of the radiative fluxes based on the predicted optical properties are mostly within \SI{2}{\flux}. The resulting radiative heating rates are also accurate, especially in the shortwave spectrum. Radiative heating rate errors may be over \SI{4}{\heat} in the longwave spectrum, mainly near the surface, but we expect these errors to decrease rapidly after adjustment of the surface and air temperatures.

The neural network-based gas optics parametrization tested in this study is up to about 4 times faster than RRTMGP, depending on network size. The larger networks achieve lower speed-ups than the small networks, but result in more accurate radiative fluxes and heating rates, clearly showing a trade-off between accuracy and computational speed. To further investigate this trade-off, we trained two additional sets of neural networks; each is tuned for the narrow range of conditions of a single LES simulation (\texttt{Cabauw}, \texttt{RCEMIP}). 
In general, these LES-tuned networks are more accurate on profiles of their respective simulations than the \texttt{NWP} networks, especially for shortwave radiation. This indicates that with LES tuning, smaller and therefore faster neural networks suffice to achieve a desired accuracy.

Given that RRTMGP uses linear interpolation from look-up tables to compute optical properties \cite{Pincus2019}, the computational efficiency of our neural network-based parametrization may be surprising. We attribute the speed-ups achieved by our parametrization to a large extent to the case-specific tuning, i.e. considering only a few gases or greatly limiting the range of atmospheric conditions (\texttt{Cabauw} and \texttt{RCEMIP} only), which reduces the problem size for which the neural networks have to be trained. Furthermore, the matrix computations required to solve the neural networks allow the use of machine-specific optimised BLAS libraries and reduces the memory use and access at the expense of floating point operations. 

The speed-ups we achieve are less than those achieved by end-to-end approaches that emulate full radiative transfer parametrizations \cite{Chevallier1998,Krasnopolsky2005,krasnopolsky2006,Krasnopolsky2010}, which may be up to 80 times faster than the original radiative transfer schemes. An advantage of our machine learning approach is that it still respects the governing radiative transfer equations, at the cost of having to perform the spectral integration by predicting optical properties and calculating fluxes for all $g$-points.
A promising future approach would be the application of machine learning to optimise the spectral integration. With such a machine learning approach the radiative transfer equations will still be solved, while the number of quadrature points may be reduced, e.g. by training neural networks to predict broadband fluxes from a small set of g-point. This would speed-up both the computations of both optical properties and the resulting radiative fluxes. 
The benefit of case-specific neural network-training also raises the question too what extent RRMTGP can be accelerated by reducing the number of input gases, which may result in smaller lookup tables and fewer computations. This was not investigated in this study, but the use of case-specific lookup tables in RRTMGP would be interesting for further studies.

\aucontribute{
MV carried out the experiments and analyses.
CvH and RP developed the ideas that led to the study. 
MV and CvH designed the study.
RS provided feedback on data generation and network design.
CvL and DP provided expert knowledge on optimising network design and hardware-specific tuning.
RP provided expert knowledge on radiative transfer.
All authors read and approved the manuscript.}

\competing{The author(s) declare that they have no competing interests.}

\funding{This study was funded by the SURF Open Innovation Lab, project no. SOIL.DL4HPC.03 and the Dutch Research Council (NWO), project no. VI.Vidi.192.068.}

\ack{The authors thank Axel Berg and the SURF Open Innovation Lab for introducing us into the world of machine learning, Peter Ukkonen for an interesting exchange of ideas, and Jordi Vil\`{a}-Guerau de Arellano for valuable discussions. RP is grateful to the conference organisers for the invitation to speak and motivation to think through the problem.}

\end{document}